\documentclass[proceedings,nohyper]{JHEP} 

\usepackage{epsfig}                   
\def\ifmath#1{\relax\ifmmode #1\else $#1$\fi}
\def\half{\ifmath{{\textstyle{1 \over 2}}}}
\def\quarter{\ifmath{{\textstyle{1 \over 4}}}}

\def\ra{\rightarrow}

\def\bs{\bigskip}
\def\ni{\noindent}
\def\lsim{\raise0.3ex\hbox{$\;<$\kern-0.75em\raise-1.1ex\hbox{$\sim\;$}}}
\def\beq{\begin{equation}}
\def\eeq{\end{equation}}
\def\ba{\begin{array}}
\def\ea{\end{array}}
\def\beqa{\begin{eqnarray}}
\def\eeqa{\end{eqnarray}}
\def\vb#1{\vbox to #1 pt{}}
\def\nn{\nonumber}

\newcommand {\ignore}[1]{}
\def\rnc#1#2#3{{\it Riv. Nuovo Cimento }{\bf #1} (#2) #3}
\conference{Trieste Meeting of the TMR Network on Physics beyond the SM}

\title{Yukawa Unification on the Bilinear R-Parity model}

\author{Jorge C. Rom\~ao\\
        Instituto Superior T\'ecnico, 
Departamento de F\'{\i}sica\\
A. Rovisco Pais 1, 1049-001 Lisboa, Portugal\\
        E-mail: \email{fromao@alfa.ist.utl.pt}}

\abstract{We discuss gauge and Yukawa unification in the context of
a supersymmetric model with bilinear R--parity violation. We show that 
this model allows $b-\tau$ Yukawa unification for
any value of $\tan\beta$ while satisfying perturbativity of the
couplings.  We also find the $t-b-\tau$ Yukawa unification 
easier to achieve than in the MSSM, occurring in a 
wider high $\tan\beta$ region. Finaly, we also discuss the compatibility
between the predicted and the measured values for $\alpha_s(M_Z)$.}


\begin{document}

\section{Introduction}

The Standard Model (SM) of particle physics is very successful in
describing the interactions of the elementary particles, except
possibly neutrinos. Although it is regarded as a good low-energy
effective theory, the SM has many theoretical problems. Its gauge
symmetry group is the direct product of three groups $SU(3)\times
SU(2)\times U(1)$ and the corresponding gauge couplings are
unrelated. It does not explain the three family structure of quarks
and leptons, and their masses are fixed by arbitrary Yukawa couplings,
with neutrinos being prevented from having mass. The Higgs sector,
responsible for the  symmetry breaking and for the fermion
masses, has not been tested experimentally and the mass of the Higgs
boson is unstable under radiative corrections.

In supersymmetry (SUSY) \cite{SUSY} the Higgs boson mass is stabilized
under radiative corrections because the loops containing standard
particles are partially canceled by the contributions from loops
containing SUSY particles. If to the Minimal
Supersymmetric Standard Model (MSSM) \cite{MSSM} we add the notion of Grand
Unified Theory (GUT), then we find that the three gauge couplings
approximately unify at a certain scale $M_{GUT}$ \cite{GUT}. Indeed, 
measurements of the gauge couplings at the CERN $e^+e^-$ collider LEP 
and neutral current data \cite{PDG} are in much better agreement with 
the MSSM--GUT with the SUSY scale $M_{SUSY}\lsim 1$ TeV 
\cite{gaugeUnif}, as compared with the SM.

Besides achieving gauge coupling unification \cite{gaugUnifRecent},
GUT theories also reduce the number of free parameters in the Yukawa
sector.  For example, in $SU(5)$ models, the bottom quark and the tau
lepton Yukawa couplings are equal at the unification scale, and the
predicted ratio $m_b/m_{\tau}$ at the weak scale agrees with
experiments. Furthermore, a relation between the top quark mass and
$\tan\beta=v_u/v_d$, the ratio between the vacuum expectation values of the
two Higgs doublets is predicted. Two solutions are possible,
characterized by low and high values of $\tan\beta$ \cite{YukUnif}.
In models with larger groups, such as $SO(10)$ and $E_6$, both the top
and bottom Yukawa couplings are unified with the tau Yukawa at the
unification scale \cite{YukUniThree}. In this case, only the large
$\tan\beta$ solution survives.

In this talk we describe some recent results \cite{YukUnifBRpV}, 
that show that the minimal extension of the MSSM--GUT
\cite{epsrad} in which R--Parity Violation (RPV) is introduced via a
bilinear term in the MSSM superpotential \cite{e3others,chaHiggsEps},
allows $b$-$\tau$ Yukawa unification for any value of
$\tan\beta$.
We also analyze the $t$-$b$-$\tau$ Yukawa unification and find that it is
easier to achieve than in the MSSM, occurring in a slightly wider high
$\tan\beta$ region. We also address the question of the compatibility
between the predicted and measured value for $\alpha_s(M_Z)$ in the
MSSM and in the bilinear RPV model.

\section{Description of the Model}

The superpotential $W$ is given by \cite{e3others,chaHiggsEps}
\begin{eqnarray}
W&\hskip -1mm=\hskip -1mm&
\varepsilon_{ab}\!\left[
 h_U^{ij}\widehat Q_i^a\widehat U_j\widehat H_u^b
\!+\! h_D^{ij}\widehat Q_i^b\widehat D_j\widehat H_d^a
\!+\! h_E^{ij}\widehat L_i^b\widehat R_j\widehat H_d^a \right.\cr
\vb{18}
&\hskip -6mm&
\left. \hskip 1cm 
-\mu\widehat H_d^a\widehat H_u^b
+\epsilon_i\widehat L_i^a\widehat H_u^b\right]
\label{superpotential}
\end{eqnarray}
where $i,j=1,2,3$ are generation indices, $a,b=1,2$ are $SU(2)$
indices. This superpotential is motivated by models of spontaneous
breaking of R--Parity~\cite{SBRpV}. Here R--Parity and lepton number
are explicitly violated by the last term in Eq.~(\ref{superpotential}).

The set of soft SUSY
breaking terms are
\begin{eqnarray}
V_{soft}&\hskip -1mm=\hskip -1mm&
M_Q^{ij2}\widetilde Q^{a*}_i\widetilde Q^a_j+M_U^{ij2}
\widetilde U^*_i\widetilde U_j+M_D^{ij2}\widetilde D^*_i
\widetilde D_j \cr
\vb{18}
&\hskip -2mm&\hskip -5mm
+M_L^{ij2}\widetilde L^{a*}_i\widetilde L^a_j
+M_R^{ij2}\widetilde R^*_i\widetilde R_j+m_{H_d}^2 H^{a*}_d H^a_d\cr
\vb{18}
&\hskip -2mm&\hskip -5mm
+m_{H_u}^2 H^{a*}_u H^a_u 
- \left[\half \sum M_i\lambda_i\lambda_i+h.c.\right]\cr
\vb{18}
&\hskip -2mm&\hskip -5mm
+\varepsilon_{ab}\left[
A_U^{ij}\widetilde Q^a_i\widetilde U_j H_u^b
+A_D^{ij}\widetilde Q^b_i\widetilde D_j H_d^a\right.\cr
\vb{18}
&\hskip -2mm& \hskip -5mm\left.
+A_E^{ij}\widetilde L^b_i\widetilde R_j H_d^a 
\!-\!B\mu H_d^a H_u^b\!+\!B_i\epsilon_i\widetilde L^a_i H_u^b\right]
\,.
\end{eqnarray}
The bilinear
R-parity violating term {\sl cannot} be eliminated by superfield
redefinition. The
reason~\cite{marco} is that the bottom Yukawa coupling, usually neglected,
plays a crucial role in splitting
the soft-breaking parameters $B$ and $B_i$ as well as the scalar
masses $m_{H_d}^2$ and $M_L^{2}$, assumed to be equal at the
unification scale.

\ni
The electroweak symmetry is broken when the VEVS of 
the two Higgs doublets $H_d$
and $H_u$, and the sneutrinos.
\begin{eqnarray}
H_d&=&{{{1\over{\sqrt{2}}}[\chi^0_d+v_d+i\varphi^0_d]}\choose{
H^-_d}} \\
H_u&=&{{H^+_u}\choose{{1\over{\sqrt{2}}}[\chi^0_u+v_u+
i\varphi^0_u]}}\\
L_i&=&{{{1\over{\sqrt{2}}}
[\tilde\nu^R_{i}+v_i+i\tilde\nu^I_{i}]}\choose{\tilde\ell^{i}}}
\end{eqnarray}
The gauge bosons $W$ and $Z$ acquire masses
\beq
m_W^2=\quarter g^2v^2 \quad ; \quad m_Z^2=\quarter(g^2+g'^2)v^2
\eeq
where
\beq
v^2\equiv v_d^2+v_u^2+v_1^2+v_2^2+v_3^2=(246 \; {\rm GeV})^2
\eeq
We introduce the
following notation in spherical coordinates:
\begin{eqnarray}
v_d&=&v\sin\theta_1\sin\theta_2\sin\theta_3\cos\beta\cr
v_u&=&v\sin\theta_1\sin\theta_2\sin\theta_3\sin\beta\cr
v_1&=&v\sin\theta_1\sin\theta_2\cos\theta_3\cr
v_2&=&v\sin\theta_1\cos\theta_2\cr
v_3&=&v\cos\theta_1\nn
\end{eqnarray}
which preserves the MSSM 
definition $\tan\beta=v_u/v_d$. The angles $\theta_i$
are equal to $\pi/2$ in the MSSM limit.

\ni
The full scalar potential may be written as

\beq
V_{total}  = \sum_i \left| { \partial W \over \partial z_i} \right|^2
+ V_D + V_{soft} + V_{RC}
\eeq
where $z_i$ denotes any one of the scalar fields in the
theory, $V_D$ are the usual $D$-terms, $V_{soft}$ the SUSY soft
breaking terms, and $V_{RC}$ are the 
one-loop radiative corrections.

\ni
In writing $V_{RC}$ we  use the diagrammatic method and find 
the minimization conditions by correcting to one--loop the tadpole
equations. 
This me\-thod has advantages with respect to the effective potential when
we calculate the one--loop corrected scalar masses. The scalar
potential contains linear terms, 
\beq
V_{linear}=t_d\sigma^0_d+t_u\sigma^0_u+t_i\tilde\nu^R_{i}
\equiv t_{\alpha}\sigma^0_{\alpha}\,,
\eeq
where we have introduced the notation
\beq
\sigma^0_{\alpha}=(\sigma^0_d,\sigma^0_u,\nu^R_1,\nu^R_2,\nu^R_3)
\eeq
and $\alpha=d,u,1,2,3$. The one loop tadpoles are
\begin{eqnarray}
t_{\alpha}&=&t^0_{\alpha} -\delta t^{\overline{MS}}_{\alpha}
+T_{\alpha}(Q)\cr
\vb{22}
&=&t^0_{\alpha} +T^{\overline{MS}} _{\alpha}(Q)
\label{tadpoles}
\end{eqnarray}
where $T^{\overline{MS}} _{\alpha}(Q)\equiv -\delta t^{\overline{MS}}_{\alpha}
+T_{\alpha}(Q)$ are the finite one--loop tadpoles.

\ni
In the following we will consider the one generation version of this
model, where only $\epsilon_3\not=0$. Then $v_1=v_2=0$ if
$\epsilon_1=\epsilon_2=0$.

\section{Main Features}

The $\epsilon$--model is a one(three) parameter(s) generalization of
the MSSM.
It can be thought as an effective model
showing the more important features of the SBRP--model~\cite{SBRpV} 
at the weak
scale. 
The mass matrices, charged and neutral currents, are similar to the
SBRP--model if we identify
\beq
\epsilon \equiv v_R h_{\nu}
\eeq
The R--Parity violating
parameters $\epsilon_3$ and $v_3$ violate tau--lepton number, inducing
a non-zero $\nu_{\tau}$ mass
$m_{\nu_{\tau}}\propto (\mu v_3+\epsilon_3v_d)^2$,
which arises due to mixing between the weak eigenstate $\nu_{\tau}$ and
the neutralinos. 
The $\nu_e$ and $\nu_{\mu}$
remain massless in first approximation.  They acquire 
masses from supersymmetric loops \cite{ralf,numass} that are typically
smaller than the tree level mass.  

The model has the MSSM as a limit. This can be illustrated in
Figure~\ref{fig1} where we show the ratio of the lightest CP-even Higgs
boson mass $m_h$ in the
$\epsilon$--model and in the MSSM  as a function of
$v_3$. 
Many other results concerning this model and the implications for
physics at the accelerators can be found in ref.~\cite{e3others,chaHiggsEps}.

\FIGURE[t]{\epsfig{file=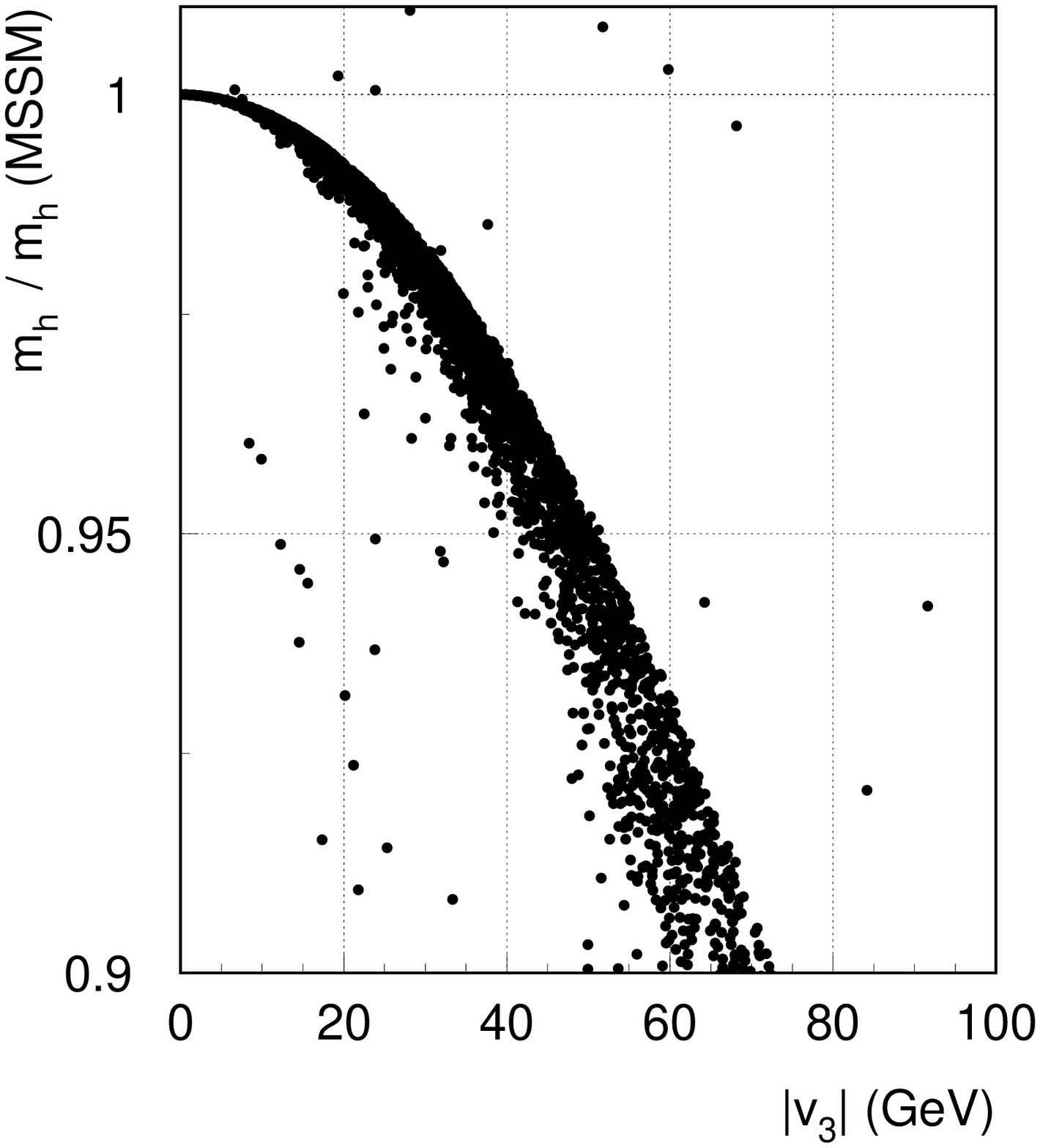,width=6.5cm}
\caption{Ratio of the 
lightest CP-even Higgs boson mass $m_h$ in the
$\epsilon$--model and in the MSSM  as a function of
$v_3$.}
\label{fig1}}

\section{Radiative Breaking}
\subsection{Radiative Breaking in the $\epsilon\!$ model: The minimal case}

At $Q = M_{GUT}$ we assume the standard minimal supergravity
unifications assumptions, 
\beqa
&&A_t = A_b = A_{\tau} \equiv A \:, \cr
&&\vb{24}
B=B_2=A-1 \:, \cr
&&\vb{24}
m_{H_1}^2 = m_{H_2}^2 = M_{L}^2 = M_{R}^2 = m_0^2 \:, \cr
&&\vb{24}
M_{Q}^2 =M_{U}^2 = M_{D}^2 = m_0^2 \:, \cr
&&\vb{24}
M_3 = M_2 = M_1 = M_{1/2} 
\eeqa
In order to determine the values of the Yukawa couplings and of the
soft breaking scalar masses at low energies we first run the RGE's from
the unification scale $M_{GUT} \sim 10^{16}$ GeV down to the weak
scale. 
We randomly give values at the unification scale
for the parameters of the theory. 
\beq
\begin{array}{ccccc}
10^{-2} & \leq &{h^2_t}_{GUT} / 4\pi & \leq&1 \cr
10^{-5} & \leq &{h^2_b}_{GUT} / 4\pi & \leq&1 \cr
-3&\leq&A/m_0&\leq&3 \cr
0&\leq&\mu^2_{GUT}/m_0^2&\leq&10 \cr
0&\leq&M_{1/2}/m_0&\leq&5 \cr
10^{-2} &\leq& {\epsilon^2_3}_{GUT}/m_0^2 &\leq& 10\cr 
\end{array}
\eeq

\ni
The value of ${h^2_{\tau}}_{GUT}/ 4 \pi$ is defined in such a way
that we get the $\tau$ lepton mass correctly. 
As the charginos mix with the tau
lepton, through a mass matrix is given by
\beq
{\bf M_C}=\left[\matrix{ 
M & {\textstyle{1\over{\sqrt{2}}}}gv_u & 0 \cr 
{\textstyle{1\over{\sqrt{2}}}}gv_d & \mu &  
-{\textstyle{1\over{\sqrt{2}}}}h_{\tau}v_3 \cr 
{\textstyle{1\over{\sqrt{2}}}}gv_3 & -\epsilon_3 & 
{\textstyle{1\over{\sqrt{2}}}}h_{\tau}v_d}\nonumber
\right] 
\eeq
Imposing that one of
the eigenvalues reproduces the observed tau mass $m_{\tau}$, $h_{\tau}$
can be solved exactly as \cite{chaHiggsEps}
\beq
h_{\tau}^2={{2m_{\tau}^2}\over{v_d}}\left[
{{1+\delta_1}\over{1+\delta_2}}
\right]\nonumber
\eeq
where the $\delta_i\,$, $i=1,2$, depend on $m_{\tau}$, on the SUSY
parameters $M,\mu,\tan\beta$ and on the R-Pari\-ty violating parameters
$\epsilon_3$ and $v_3$. 
It can be shown \cite{chaHiggsEps} that
\beq
\lim_{\epsilon_3 \ra 0} \delta_i = 0
\eeq

\ni
After running the RGE we have a 
complete set of parameters, Yukawa couplings and soft-breaking masses 
$m^2_i(RGE)$ to study the minimization. To do this we use the
following  method \cite{epsrad}:

\begin{enumerate}

\item
We start with random values for $h_t$ and $h_b$ at $M_{GUT}$. 
The value of $h_{\tau}$ at $M_{GUT}$
is fixed in order to get the correct $\tau$ mass.

\item
The value of $v_d$ is determined from $m_{b}=h_b v_d/ \sqrt{2}$ for
$m_{b}=2.8$ GeV (running $b$ mass at $m_Z$). 

\item
The value of $v_u$ is determined from $m_{t}=h_t v_u/ \sqrt{2}$ for
$m_{t}=176 \pm 5$ GeV. If 
\beq
\hskip -0.5mm
v_d^2+v_u^2 \!>\! v^2=\frac{4}{g^2}\, m^2_W = (246 \hbox{ GeV})^2
\eeq
then we go back and choose another starting point.
The value of $v_3$ is then obtained from
\beq
v_3=\pm\, \sqrt{\frac{4}{g^2}\, m^2_W -v_1^2 -v_2^2}
\eeq
\end{enumerate}

\ni
We see that the freedom in $h_{t}$ and $h_{b}$ at $M_{GUT}$ can be
translated into the freedom in the mixing angles $\beta$ and
$\theta$. Comparing, at this point, with the MSSM we have one extra
parameter $\theta$. We will discuss this in more detail below. In 
the MSSM we would have $\theta=\pi/2$.
After doing this, for each point in parameter space, we solve the extremum
equations, for the soft breaking
masses, which we now call $m^2_i$ ($i=H_1,H_2,L$). 
Then we calculate numerically the eigenvalues for the real and
imaginary part of the neutral scalar mass-squared matrix. If they are
all positive, except for the Goldstone boson, the point is a good one. 
If not, we go back to the next random value. 
As before, we end up
with a set of solutions for which
the $m^2_i$ obtained from the minimization
of the potential differ from those obtained from the RGE, which we
call  $m^2_i(RGE)$. 
Our  goal is to find solutions that obey
\beq
m^2_i=m^2_i(RGE) \quad \forall i
\eeq
To do that we define a function
\beq
\eta= max \left( \frac{m^2_i}{m^2_i(RGE)},\frac{m^2_i(RGE)}{m^2_i}
\right) \quad \forall i 
\eeq
We see that we have always
\beq
\eta \ge 1
\eeq
and use {\tt MINUIT} to minimize $\eta$. We have shown \cite{epsrad}
that it is easy to get solutions for this problem. 

Before we finish this section 
let us discuss the counting of free
parameters.  In the minimal N=1 supergravity unified
version of the MSSM this is shown in Table~\ref{table1}. The counting
for the $\epsilon$--model is presented in Table~\ref{table2}.  
Finally, we note that in either case, the sign of the mixing parameter
$\mu$ is physical and has to be taken into account.

\TABLE{
\begin{tabular}{ccc}\hline
Parameters
\hskip -8pt&\hskip -8pt 
Conditions 
\hskip -8pt&\hskip -8pt 
Free Parameters\hskip -8pt \cr \hline
$h_t$, $h_b$, $h_{\tau}$
\hskip -8pt&\hskip -8pt
$m_W$, $m_t$
\hskip -8pt&\hskip -8pt $\tan \beta$ \cr 
$v_d$, $v_u$,$M_{1/2}$ 
\hskip -8pt&\hskip -8pt
$m_b$, $m_{\tau}$ 
\hskip -8pt&\hskip -8pt 2 Extra \cr 
$m_0$, $A$, $\mu$
\hskip -8pt&\hskip -8pt
$t_i=0$, $i=1,2$
\hskip -8pt& \hskip -8pt({\it e.g.} $m_h$, $m_A$)\cr \hline
Total = 9\hskip -8pt&\hskip -8ptTotal = 6 \hskip -8pt&\hskip -8pt
Total = 3\cr\hline
\end{tabular}
\caption{Counting of free parameters in N=1 supergravity MSSM.}
\label{table1}}
\TABLE{
\begin{tabular}{ccc}\hline
Parameters \hskip -8pt & \hskip -8pt 
Conditions \hskip -8pt & \hskip -8pt Free Parameters \cr \hline
$h_t$, $h_b$, $h_{\tau}$
\hskip -8pt & \hskip -8pt$m_W$, $m_t$ \hskip -8pt & \hskip -8pt 
$\tan\beta$, $\epsilon_i$ \cr 
$v_d$, $v_u$, $M_{1/2}$
\hskip -8pt & \hskip -8pt$m_b$, $m_{\tau}$ \hskip -8pt & \hskip -8pt \cr 
$m_0$,$A$, $\mu$
\hskip -8pt & \hskip -8pt$t_i=0$\hskip -8pt & \hskip -8pt 2 Extra \cr 
$v_i$, $\epsilon_i$
\hskip -8pt & \hskip -8pt($i=1,\ldots,5$)
\hskip -8pt & \hskip -8pt ({\it e.g.} $m_h$, $m_A$)\cr \hline
Total = 15\hskip -8pt & \hskip -8ptTotal = 9 
\hskip -8pt & \hskip -8ptTotal = 6\cr\hline
\end{tabular}
\caption{Counting of free parameters in our model.}
\label{table2}}

\subsection{Yukawa Unification in the $\epsilon$ model: I Motivation}

There is a strong motivation to consider GUT theories where {\it
both} gauge and Yukawa unification can achieved. This is because 
besides achieving gauge coupling unification,
GUT theories can also reduce the number of free parameters in the Yukawa
sector and this is normally a desirable feature. The situation with
respect to GUT theories that embed the MSSM can be summarized as
follows \cite{YukUnif,YukUniThree}:

\begin{itemize}
\item
In $SU(5)$ models, $h_b=h_{\tau}$ at $M_{GUT}$. The
predicted ratio $m_b/m_{\tau}$ at $M_{WEAK}$ agrees with
experiments. 
\bs

\item
A relation between $m_{top}$ and $\tan\beta$ is predicted. 
Two solutions are possible: low and high $\tan\beta$ .
\bs

\item
In $SO(10)$ and $E_6$ models $h_t=h_b=h_{\tau}$ at $M_{GUT}$.
In this case, only the large $\tan\beta$ solution survives.
\bs

\item
Recent global fits of low energy data (the lightest Higgs 
mass and $B(b\rightarrow s\gamma)$) to MSSM
show that it is hard to reconcile these constraints
with the large $\tan\beta$ solution.  Also the low $\tan\beta$ solution
with $\mu<0$ is also disfavored. 

\end{itemize}

In the following sections we will
show \cite{YukUnifBRpV} 
that the $\epsilon$--model allows $b-\tau$ Yukawa unification for
any value of $\tan\beta$ and satisfying perturbativity of the
couplings.  We also find the $t-b-\tau$ Yukawa unification 
easier to achieve than in the MSSM, occurring in a 
wider high $\tan\beta$ region.

\subsection{Yukawa Unification in the $\epsilon$ model: II The Method}

As before $h_{\tau}$ can be solved exactly 
\beq
h_{\tau}^2={{2m_{\tau}^2}\over{v_d}}\left[
{{1+\delta_1}\over{1+\delta_2}}
\right]
\eeq
where the $\delta_i\,$, $i=1,2$, depend on $m_{\tau}$, on the SUSY
parameters $M,\mu,\tan\!\beta$ and on the R-parity violating parameters
$\epsilon_3$ and $v_3$.
Also $h_t $ and $h_b$
are related to $m_t$ and $m_b$
\beq
m_t = h_t \frac{v}{\sqrt2} \sin \beta \sin \theta\,, \: \: \: \: \:
m_b = h_b \frac{v}{\sqrt2} \cos \beta \sin \theta 
\eeq
where
\beq
v=2m_W/g \ ; \ \tan\beta=v_u/v_d \ ; \
\cos\theta=v_3/v
\eeq

\ni 
In our approach we divide the evolution in three ranges:

\begin{enumerate}

\item
$m_{Z} \ra m_t$\\
 We use running fermion masses and gauge 
couplings.

\item
$m_t \ra M_{SUSY}$ \\
We use the two-loop SM RGE's including the quartic Higgs coupling $\lambda$.

\item
$M_{SUSY} \ra M_{GUT}$\\
We use the two-loop RGE's. 

\end{enumerate}

\ni
Using a top $\ra$ bottom 
approach we randomly vary the unification scale
$M_{GUT}$ and the unified coupling $\alpha_{GUT}$ looking for
solutions compatible with the low energy data \cite{LEPinternal}
\beqa
&&\alpha^{-1}_{em}(m_Z) = 128.896 \pm0.090\cr
&&\vb{18}
\sin^2\theta_w(m_Z) =
0.2322 \pm 0.0010\cr
&&\vb{18}
\alpha_s(m_Z)=0.118 \pm 0.003
\eeqa
We get a region centered around 
\beq
M_{GUT} \approx
2.3 \times10^{16} GeV \ ; \
{\alpha_{GUT}}^{-1} \approx 24.5
\eeq

\ni
Next we use a bottom $\ra$ top approach to 
study the unification of Yukawa couplings using two-loop
RGEs. We take \cite{LEPinternal}
\beqa
&&m_W = 80.41 \pm 0.09\ GeV\cr
&&\vb{18}
m_{\tau}=1777.0 \pm 0.3 \ MeV \cr
&&\vb{18}
m_b(m_b) = 4.1\ \hbox{to}\ 4.5\ GeV 
\eeqa
We calculate the running masses 
\beqa
m_{\tau}(m_t) &=& \eta_{\tau}^{-1} m_{\tau}(m_{\tau})\cr
\vb{18}
m_b(m_t)&=&\eta_b^{-1}m_b(m_b)
\eeqa
where $\eta_{\tau}$ and $\eta_b$
include three--loop order QCD and one--loop order QED \cite{alf3}. 
At the scale $Q=m_t$ we keep as a free parameter the running top quark
mass $m_t(m_t)$ and vary randomly the SM quartic Higgs coupling
$\lambda$. 
In solving the RG equations we take the following boundary conditions:

\begin{enumerate}
\item
At scale $Q=m_t$
\beq
\lambda_i^2(m_t)=2m_i^2(m_t)/v^2 \ ; \
i=t,b,\tau
\eeq

\item
At scale $Q=M_{SUSY}$
\beqa
\lambda_t(M_{SUSY}^-)&=&h_t (M_{SUSY}^+) \sin\beta\sin\theta \cr
\vb{18}
\lambda_b(M_{SUSY}^-)&=&h_b (M_{SUSY}^+) \cos\beta\sin\theta \cr
\vb{18}
\lambda_{\tau}(M_{SUSY}^-)&=&h_{\tau} (M_{SUSY}^+)
\cos\beta\sin\theta\cr
&&\vb{20}
\times
\sqrt{{1+\delta_2}\over{1+\delta_1}}
\eeqa
where $h_i$ denote the Yukawa couplings of our model and $\lambda_i$
those of the SM.
The boundary condition for the quartic Higgs coupling is
\beqa
\lambda(M_{SUSY}^-) &\hskip -2mm=\hskip -2mm& 
\quarter\!
\Big[(g^2(M_{SUSY}^+)\!+\!g'^2(M_{SUSY}^+) \Big] \cr
&&\vb{18}
(\cos2\beta\sin^2\theta+
\cos^2\theta)^2 
\eeqa
The MSSM limit is obtained setting $\theta \to \pi/2$ i.e. $v_3=0$.

\end{enumerate}

\ni
Before we close this section we give some details of the calculation.
At the scale $Q=M_{SUSY}$ we vary randomly the SUSY parameters $M$,
$\mu$ and $\tan\beta$, as well as the R--Parity violating parameter
$\epsilon_3$. 
The parameter $v_3=v\cos\theta$ is calculated from the boundary conditions.
Since $\lambda$ (or equivalently the SM Higgs 
mass $m_H^2=2\lambda v^2$) is varied randomly, in practice we also scan
over $\theta$. 
This way, we consider all possible initial conditions
for the RGEs at $Q=M_{SUSY}$, and evolve them up to the unification
scale $Q=M_{GUT}$.  
The solutions that satisfy $b-\tau$ unification
are kept.

\subsection{Yukawa Unification in the $\epsilon$ model: III Results
and Discussion}

\FIGURE[t]{\epsfig{file=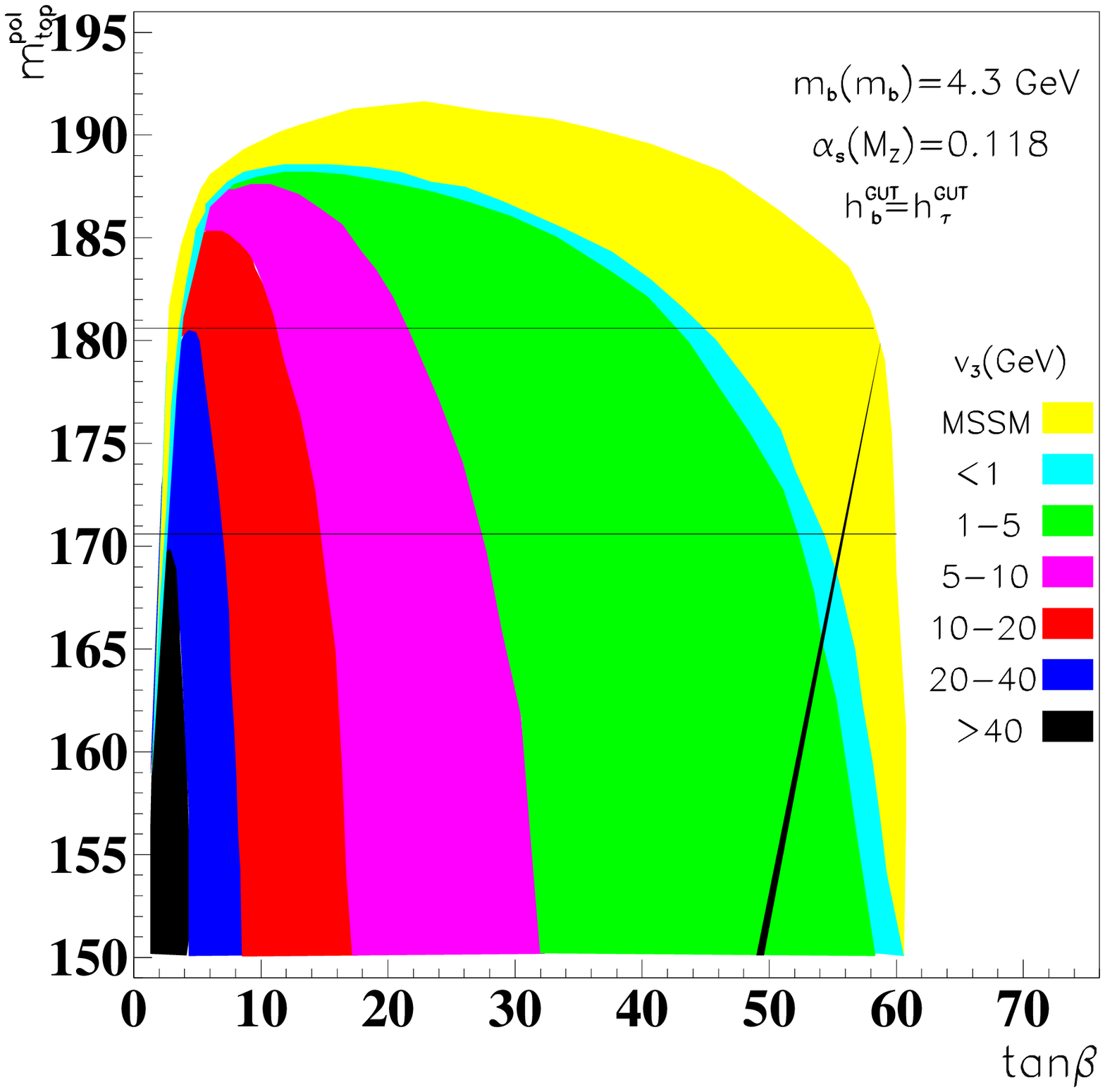,width=7cm}
\caption{Top quark mass as a function of $\tan\beta$ for different
values of the R--Parity violating parameter $v_3$. Bottom quark and
tau lepton Yukawa couplings are unified at $M_{GUT}$. The horizontal
lines correspond to the $1\sigma$ experimental $m_t$
determination. Points with $t-b-\tau$ unification lie in the diagonal
band at high $\tan\beta$ values. We have taken $M_{SUSY}=m_t$.}
\label{fig2}}

The results are summarized in Figure~\ref{fig2} where we present the
top quark mass as a function of $\tan\beta$ for different
values of the R--Parity violating parameter $v_3$. Bottom quark and
tau lepton Yukawa couplings are unified at $M_{GUT}$. The horizontal
lines correspond to the $1\sigma$ experimental $m_t$
determination. Points with $t-b-\tau$ unification lie in the diagonal
band at high $\tan\beta$ values. We have taken $M_{SUSY}=m_t$.
The dependence of our results on $\alpha_s$ and $m_b$
is totally analogous to what happens in
the MSSM. The upper bound on $\tan\beta$, which
is $\tan\beta\lsim 61$ for $\alpha_s=0.118$, increases with $\alpha_s$
and becomes $\tan\beta\lsim 63$ (59) for $\alpha_s=0.122$ (0.114).
The top mass value for which unification is achieved for any
$\tan\beta$ value within the perturbative region increases with
$\alpha_s$, as in the MSSM.  
As for the dependence on $m_b$, if we consider $m_b(m_b)=4.1$ (4.5)
GeV then the upper bound of this parameter is given by $\tan\beta\lsim
64$ (58). In addition, the MSSM region is narrower (wider) at high
$\tan\beta$ compared with the $m_b(m_b)=4.3$ GeV case.
The line at high $\tan\beta$ values corresponds
to points where $t-b-\tau$ unification is achieved. Since the region
with $|v_3|<5$ GeV overlaps with the MSSM region, it follows that
$t-b-\tau$ unification is possible in this model for values of $|v_3|$
up to about 5 GeV, instead of 50 GeV or so, which holds in the case of
bottom-tau unification.

\section{On $\alpha_3(M_Z)$ versus $\sin^2 \theta_W (M_Z)$}

Recent studies \cite{MSSMalfas} of gauge coupling 
unification in the context of minimal
R--Parity conserving supergravity (SUGRA) agree that using
the experimental values for the electromagnetic coupling and the weak
mixing angle, the prediction obtained for $\alpha_s(M_Z) \sim 0.129
\pm 0.010$ is about 2$\sigma$ larger than indicated by the most recent
world average value $\alpha_s(M_Z)^{W.A}=0.1189 \pm 0.0015$
\cite{alphasWA}. 

We have re-considered the $\alpha_s$ prediction in the context of the
model with bilinear breaking of R--Parity. We have shown
\cite{RPValfas}, that in this simplest SUGRA R--Parity breaking model,
with the same particle content as the MSSM, there appears an
additional negative contribution to $\alpha_s$, which can bring the
theoretical prediction closer to the experimental world average. This
additional contribution comes from two--loop b--quark Yukawa effects
on the renormalization group equations for $\alpha_s$. Moreover we
have shown that this contribution is typically correlated to the
tau--neutrino mass which is induced by R--Parity breaking and which
controls the R-Parity violating effects. We found that it is possible
to get a 5\% effect on $\alpha_s(M_Z)$ even for light $\nu_{\tau}$
masses. The results are summarized in Figure~\ref{fig3} where we
present the situation for the MSSM and in Figure~\ref{fig4} where the
results for the bilinear R--Parity breaking model are shown.

\FIGURE[t]{\epsfig{file=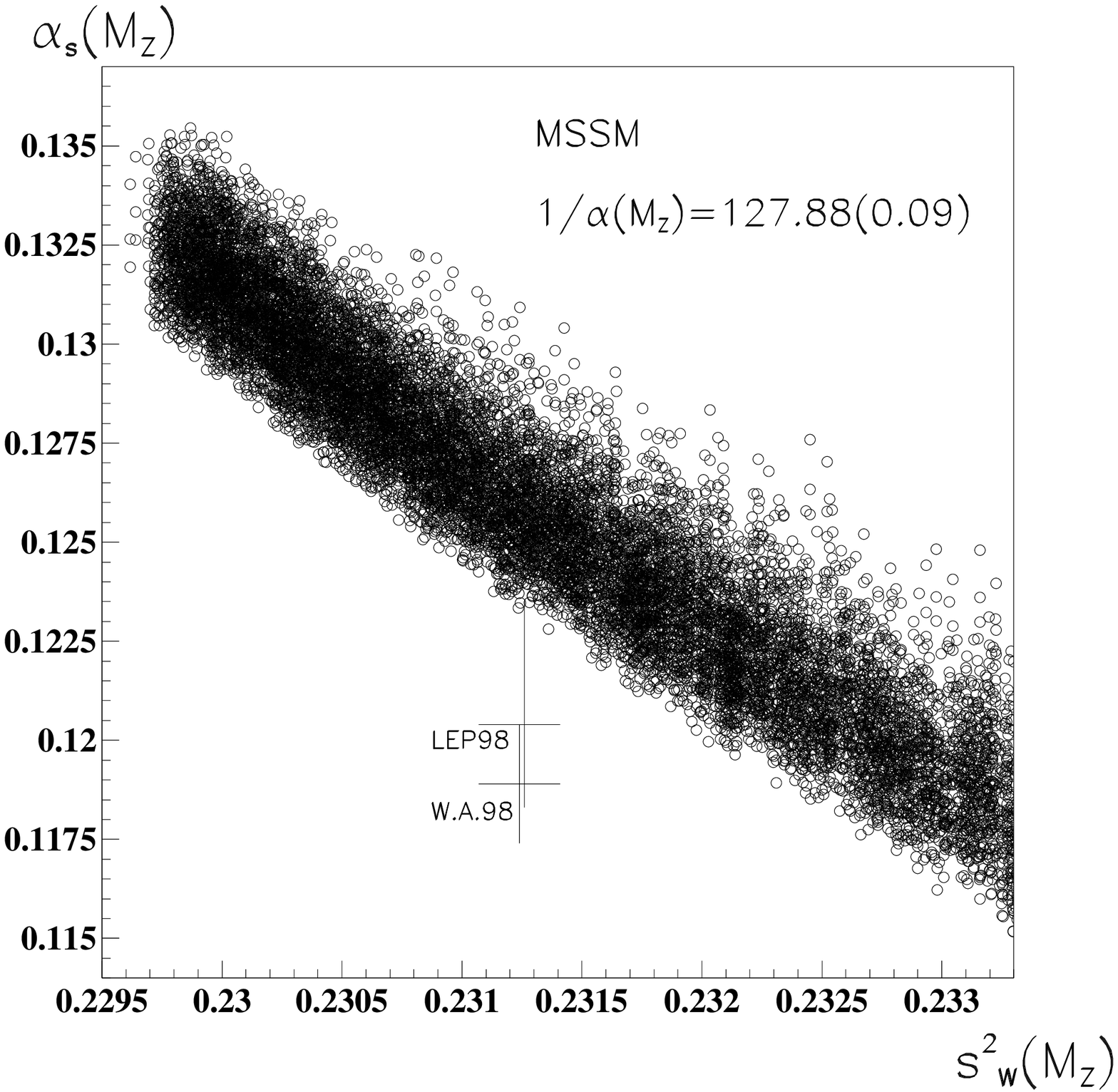,width=7cm}\caption{$\alpha_s(M_Z)$
versus $\hat s_Z$ for the MSSM.}\label{fig3}}

\FIGURE[t]{\epsfig{file=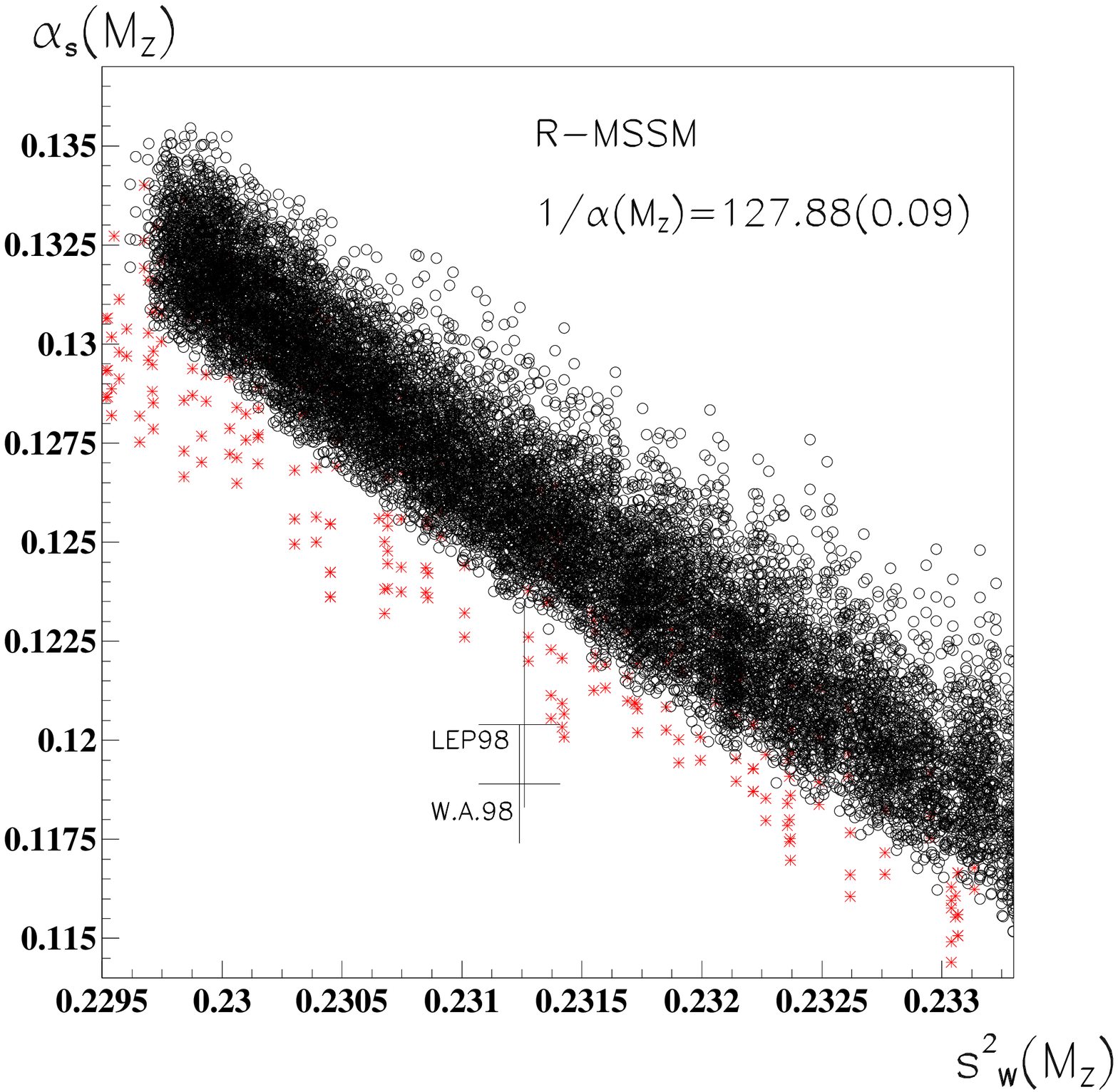,width=7cm}\caption{$\alpha_s(M_Z)$
versus $\hat s_Z$ for the bilinear $\not{\!\!\!\!R}_p$ model.}\label{fig4}}

\section{Conclusions}

The bilinear R--Parity model is a minimal extension of the MSSM with many 
new features, among which the possibility of having masses for the neutrinos.
We have shown that it is possible to incorporate these models in 
a N=1 SUGRA scenario, where the number of free parameters is reduced. 
In these so--called {\it radiative breaking} scenarios we showed that 
this model allows $b-\tau$ Yukawa unification for
any value of $\tan\beta$ while satisfying perturbativity of the
couplings.  We also find the $t-b-\tau$ Yukawa unification 
easier to achieve than in the MSSM, occurring in a 
wider high $\tan\beta$ region.
By performing a full two--loop calculation~\cite{RPValfas} 
we also have shown that in this model there appears an
additional negative contribution to $\alpha_s$, which can bring the
theoretical prediction closer to the experimental world average.
Although we presented here only the one generation example, we have
achieved also the above results in the full three generation case. In
this situation we can get at one--loop 
non zero values for the masses of the two
lightest neutrinos which very interesting in the
context of solving the solar and atmospheric neutrino 
problems~\cite{numass}.

\section*{Acknowledgements} 

This work was supported in part by the TMR network grant
ERBFMRXCT960090 of the European Union.


\begin{thebibliography}{99}

\bibitem{SUSY}
Yu.A. Gol'fand and E.P. Likhtman, \jetpl{13}{1971}{323};
D.V. Volkov and V.P. Akulov, \jetpl{16}{1972}{438};
J. Wess and B. Zumino, \npb{70}{1974}{39}.

\bibitem{MSSM}
H.P. Nilles, \prep{110}{1984}{1};
H.E. Haber and G.L. Kane, \prep{117}{1985}{75};
R. Barbieri, \rnc{11}{1988}{1}.

\bibitem{GUT}
S. Dimopoulos, S. Raby, and F. Wilczek, \prd{24}{1981}{1681};
S. Dimopoulos and H. Georgi, \npb{193}{1981}{150};
L. Iba\~nez and G.G. Ross, \plb{105}{1981}{439};
M.B. Einhorn and D.R.T. Jones, \npb{196}{1982}{475};
W.J. Marciano and G. Senjanovic, \prd{25}{1982}{3092}.

\bibitem{PDG}
Review of Particle Properties, \prd{54}{1996}{1}.


\bibitem{gaugeUnif}
U. Amaldi, W. de Boer, and H. Furstenau, \plb{260}{1991}{447};
J. Ellis, S. Kelley, and D.V. Nanopoulos, \plb{260}{1991}{131};
P. Langacker and M. Luo, \prd{44}{1991}{817};
C. Giunti, C.W. Kim and U.W. Lee, \mpla{6}{1991}{1745}.


\bibitem{gaugUnifRecent}
For recent studies see P. Langacker and N. Polonsky, \prd{47}{1993}{4028};
P.H. Chankowski, Z. Pluciennik, and S. Pokorski, \npb{439}{1995}{23};
P.H. Chankowski, Z. Pluciennik, S. Pokorski, and C.E. Vayonakis, 
\plb{358}{1995}{264}.


\bibitem{YukUnif}
V. Barger, M.S. Berger, and P. Ohmann, \prd{47}{1993}{1093};
M. Carena, S. Pokorski, and C.E.M. Wagner, \npb{406}{1993}{59};
R. Hempfling, \prd{49}{1994}{6168}.

\bibitem{YukUniThree}
L.J. Hall, R. Rattazzi, and U. Sarid, \prd{50}{1994}{7048};
M. Carena, M. Olechowski, S. Pokorski, and C.E.M. Wagner, \npb{426}{1994}{269}.

\bibitem{YukUnifBRpV}
M.A. D\'{\i}az, J. Ferrandis, J.C. Rom\~ao, and J.W.F. Valle, 
\plb{453}{1999}{263}.

\bibitem{epsrad}
M.A. D\'{\i}az, J.C. Rom\~ao, and J.W.F. Valle, \npb{524}{1998}{23};
M.A. D\'{\i}az, talk given at International Europhysics Conference on
High-Energy Physics, Jerusalem, Israel, 19-26 Aug 1997,
\hepph{9712213}; J.C. Rom\~ao, talk given at International
Workshop on Physics Beyond the Standard Model: From Theory to
Experiment (Valencia 97), Valencia, Spain, 13-17 Oct 1997,
\hepph{9712362};
J.W.F. Valle, review talk given at the Workshop on Physics Beyond
the Standard Model: Beyond the Desert: Accelerator and Nonaccelerator
Approaches, Tegernsee, Germany, 8-14 Jun 1997, \hepph{9712277}.

\bibitem{e3others}
F. de Campos, M.A. Garc{\'\i}a-Jare\~no, A.S. Joshipura, J. Rosiek,
and J. W. F. Valle, \npb{451}{1995}{3};T. Banks,
Y. Grossman, E. Nardi, and Y. Nir, \prd{52}{1995}{5319};
A. S. Joshipura and M.Nowakowski, \prd{51}{1995}{2421};
R. Hempfling, \npb{478}{1996}{3};
F. Vissani and A.Yu. Smirnov, \npb{460}{1996}{37};
H. P. Nilles and N. Polonsky, \npb{484}{1997}{33};
B. de Carlos, P. L. White, \prd{55}{1997}{4222};
S. Roy and B. Mukhopadhyaya, \prd{55}{1997}{7020}.


\bibitem{chaHiggsEps}
A. Akeroyd, M.A. D\'{\i}az, J. Ferrandis, M.A. Garcia--Jare\~no, and
J.W.F. Valle, \npb{529}{1998}{3}.


\bibitem{SBRpV}
A. Masiero and J.W.F. Valle, \plb{251}{1990}{273};
M.C. Gonzalez-Garcia, J.W.F. Valle, \npb{355}{1991}{330};
J.C. Rom\~ao, C.A. Santos, J.W.F. Valle, \plb{288}{1992}{311};
J.C. Rom\~ao, A. Ioannissyan and J.W.F. Valle, \prd{55}{1997}{427}.

\bibitem{marco}
M.A. D\'{\i}az, talk given at International
Workshop on Physics Beyond the Standard Model: From Theory to
Experiment (Valencia 97), Valencia, Spain, 13-17 Oct 1997,
\hepph{9802407}.

\bibitem{ralf}
R. Hempfling, \npb{478}{1996}{3}.

\bibitem{numass}
M.A. D\'{\i}az, M. Hirsch, W. Porod, J.C. Rom\~ao and J.W.F. Valle in
preparation. See also J.C. Rom\~ao talk at the International Workshop
on {\sl Particles in Astrophysics and Cosmology: 
From Theory to Observation}, Val\`encia, Spain, 3-8 May 1999. 

\bibitem{LEPinternal}
``A Combination of Preliminary Electroweak Measurements and Constraints
on the Standard Model'', CERN internal note, LEPEWWG/97-02, Aug. 1997.

\bibitem{alf3}
O. V. Tarasov, A. A. Vladimirov, and A. Y. Zharkov, \plb{93}{1980}{429};
S.G. Gorishny, A.L. Kateav, and S.A. Larin, \yf{40}{1984}{517}
[\sjnp{40}{1984}{329}];
S.G. Gorishny {\it et. al.}, \mpla{5}{1990}{2703}.

\bibitem{MSSMalfas}
P. Langacker and N. Polonsky, \prd{47}{1993}{4028};
P. Langacker and N. Polonsky, \prd{52}{1995}{3081};
M. Carena, S. Pokorski and C.E.M. Wagner, \npb{406}{1993}{59}.

\bibitem{alphasWA}
C. Caso {\it et. al.} \epjc{3}{1998}{1}. 


\bibitem{RPValfas}
M.A.D\'{\i}az, J. Ferrandis, J.C. Rom\~ao and J.W.F. Valle,
\hepph{9906343}.



\end{thebibliography}
\end{document}